\documentclass[runningheads]{llncs}
\usepackage[T1]{fontenc}
\usepackage[utf8]{inputenc}
\usepackage{graphicx}
\usepackage{hyperref}
\hypersetup{hidelinks}
\usepackage{booktabs}
\usepackage[frozencache,cachedir=.]{minted}
\usepackage{tabularx}
\usepackage{stmaryrd}
\usepackage[inline]{enumitem}

\usepackage{amsmath}
\usepackage{pict2e}

\newcommand{\lbparen}{%
  \mathopen{%
    \sbox0{$()$}%
    \setlength{\unitlength}{\dimexpr\ht0+\dp0}%
    \raisebox{-\dp0}{%
      \begin{picture}(.32,1)
      \linethickness{\fontdimen8\textfont3}
      \roundcap
      \put(0,0){\raisebox{\depth}{$($}}
      \polyline(0.32,0)(0,0)(0,1)(0.32,1)
      \end{picture}%
    }%
  }%
}

\newcommand{\rbparen}{%
  \mathclose{%
    \sbox0{$()$}%
    \setlength{\unitlength}{\dimexpr\ht0+\dp0}%
    \raisebox{-\dp0}{%
      \begin{picture}(.32,1)
      \linethickness{\fontdimen8\textfont3}
      \roundcap
      \put(-0.08,0){\raisebox{\depth}{$)$}}
      \polyline(0,0)(0.32,0)(0.32,1)(0,1)
      \end{picture}%
    }%
  }%
}

\newcommand{\code}[1]{\texttt{#1}}
\newcommand{\ie}{\emph{i.e.},\ }
\usepackage{color}

\begin{document}
\sloppy

\title{Going Bananas! - Unfolding Program Synthesis with Origami}
\titlerunning{Going Bananas!}
%
\author{
Matheus Campos Fernandes\inst{1}\orcidID{0000-0002-5118-2399} 
\and
Fabrício Olivetti de França\inst{1}\orcidID{0000-0002-2741-8736} 
\and
Emilio Francesquini\inst{1}\orcidID{0000-0002-5374-2521}}
\authorrunning{Fernandes et al.}
%
\institute{Federal University of ABC (UFABC), Santo André, São Paulo, Brazil\\
\email{\{fernandes.matheus,folivetti,e.francesquini\}@ufabc.edu.br}}
\maketitle              
\begin{abstract}
Automatically creating a computer program using input-output examples can be a challenging task, especially when trying to synthesize computer programs that require loops or recursion.
Even though the use of recursion can make the algorithmic description more succinct and declarative, this concept creates additional barriers to program synthesis algorithms such as the creation and the (tentative) evaluation of non-terminating programs. One reason is that the recursive function must define how to traverse (or generate) the data structure and, at the same time, how to process it. 
In functional programming, the concept of recursion schemes decouples these two tasks by putting a major focus on the latter. This can also help to avoid some of the pitfalls of recursive functions during program synthesis, as argued in a previous work where we introduced the Origami technique. In our previous paper, we showed how this technique was effective in finding solutions for programs that require folding lists. 
In this work, we incorporate other recursion schemes into Origami, such as accumulated folding, unfolding, and the combination of unfolding and folding.
We evaluated Origami on the 29 problems of the standard General Program Synthesis Benchmark Suite 1, obtaining favorable results against other well-known algorithms.
Overall, Origami achieves the best result in $25\%$ more problems than its predecessor (HOTGP) and an even higher increase when compared to other approaches. Not only that, but it can also consistently find a solution to problems that many algorithms report a low success rate.

\end{abstract}
\setlength{\textfloatsep}{2em}

\section{Introduction}
\label{sec:intro}

\emph{Program Synthesis} (PS) is the task of automatically generating a computer program written in the programming language of choice using some form of specification~\cite{gulwani2010dimensions}.
This task can be framed as a search problem, in which we explore the space of possible programs under a set of constraints, usually by a grammar and a maximum program size, that remove irrelevant or undesirable solutions from the search space.
From this point of view, this process depends on three factors: the format of that specification, the constraints of the search space, and the actual search algorithm~\cite{gulwani2010dimensions}.

A convenient format for the specification is a set of input-output examples that demonstrate the expected outputs for several different input cases.
In this particular case, the task is called \emph{Inductive Synthesis} or \emph{Programming-by-Example} (PBE).
The main advantage of this approach is that sets of examples are easy to create, and often do not require deep knowledge of the problem in question.
However, the lack of corner or special cases in the set allows for the generation of programs that do not follow the original intent of the user.

Among the vast selection of possible search algorithms, we highlight \emph{Genetic Programming} (GP)~\cite{koza1992genetic}. 
GP is a search algorithm that tries to balance the exploration and exploitation of the search space using recombination and perturbation of a bag of solutions. 
It employs selective pressure at every step, inspired by the evolution of species, to favor the fittest solutions when replicating and applying such operators.
Some benchmark problems extracted from common programming tasks have been successfully solved by recent variations of the original GP algorithm, such as PushGP~\cite{helmuth2018program}, CBGP~\cite{pantridge2020code}, GE~\cite{o2001grammatical}, G3P~\cite{g3p} and HOTGP~\cite{hotgp}.

The representation of the solutions and the imposed constraints play a major role in search algorithms, as they describe the navigability in this space and the coverage of possible programs.
For example, the Push language~\cite{spector2002genetic} is a stack-based programming language in which the operations are executed sequentially storing and retrieving values from the stack corresponding to the types of the operation.
As such, it provides additional flexibility to the computer programs as a certain operation is not constrained by the type of the previous one. 
This increases the possibilities of navigation of the search space. 
On the other hand, in Grammatical Evolution~\cite{o2001grammatical} the search space is defined by the grammar being evolved. 
This can limit the size of the search space and ensure that each step of the generated program is valid.
Likewise, type-safe representations, such as in CBGP~\cite{pantridge2020code} and HOTGP~\cite{hotgp}, carry at each node the information of the input and output types, constraining the space to valid programs (\textit{i.e.}, that execute without errors) and solutions that follow the type specification.


A common challenge to GP algorithms is how to handle recursion or loops. The problem here lies in the fact that it is possible to generate an unnecessarily long or even unbounded recursion/loop.
This is alleviated by the use of programming patterns that hide the loop or recursion behind a declarative language.
Popular examples of such patterns are the \code{map}, \code{filter}, and \code{fold} functions.

In particular, \emph{fold} describes a recursive pattern capable of expressing recursive algorithms that traverse a structure aggregating the partial results.
Examples of the use of this pattern are \emph{sum}, \emph{product}, and even \emph{insertion sort}.
A complementary pattern that describes the recursion that generates or builds a structure is known as \emph{unfold}.
For example, \texttt{unfold} can  be used to generate the list of Fibonacci numbers, or reversing a list.

Although many algorithms can be described using folds or unfolds, this approach has some limitations such as working only on lists, or the inability to store partial results.
A more general set of recursive patterns that involves folding and unfolding is known as \emph{Recursion Schemes}~\cite{meijer1991functional}.
The Recursion Schemes extend the common folding and unfolding operations to work on any inductive type and include additional mechanisms to handle a wider variety of recursive function patterns%
\footnote{The authors represent these recursive functions using bananas, lenses, envelopes, and barbed wires~\cite{meijer1991functional}.}%
.
These recursion schemes can be used to guide the synthesis of recursive programs, as their structure is well-defined and can be used as scaffolding, with variations only in certain parts of the program.

Origami~\cite{origami} is an algorithm capable of synthesizing typed, pure, functional programs that support different Recursion Schemes.
In that work, a proof-of-concept implementation of Origami was presented, capable of partially synthesizing one single morphism. This implementation was evaluated, yielding promising preliminary results.
In this paper, we provide and evaluate the first complete implementation of Origami, following the description given in the original paper.
This represents a meaningful step in assessing the effectiveness of integrating Recursion Schemes into Program Synthesis.
We evaluate the performance of Origami using the problems described in the General Program Synthesis Benchmark Suite 1 (PSB1)~\cite{psb1}, comparing the obtained results to well-known GP-based program synthesis algorithms. 
When comparing the success rate of each problem, Origami achieves the best result in $25\%$ more problems than its predecessor (HOTGP) and an even higher increase when compared to other approaches. 
Additionally, Origami was capable of achieving a high success rate (higher than $70\%$) in problems that most algorithms achieve less than $50\%$, such as \textit{scrabble-score}, \textit{grade}, \textit{for-loop-index}, and \textit{super-anagrams}.

The remainder of this text is organized as follows. In~\autoref{sec:related} we conduct a brief literature review of Functional Program Synthesis and Recursion Schemes. In~\autoref{sec:methods} we present Origami and outline details of our implementation. In~\autoref{sec:results} we show and analyze the results, comparing Origami to HOTGP and other well-known methods. Finally, in~\autoref{sec:conclusion} we give some final observations about Origami and describe future work.

\section{Related Work}
\label{sec:related}

HOTGP~\cite{hotgp} is a GP algorithm that synthesizes pure, typed, and functional programs. 
Its approach to recursion includes support for higher-order functions, $\lambda$-functions, and parametric polymorphism. 
Experiments conducted on PSB1 showed that HOTGP is capable of synthesizing correct programs more frequently than any other of the evaluated algorithms, on average.

In~\cite{swan2019stochastic}, the authors presented the first algorithm for synthesizing programs that exploit Recursion Schemes. 
Their work focuses only on catamorphisms over natural numbers using Peano representation (\ie the inductive type of natural numbers). The authors evaluate their approach with variations of the Fibonacci sequence, successfully obtaining the correct programs.

Our approach, Origami, was originally proposed in~\cite{origami} where we evaluated the feasibility of using Recursion Schemes to synthesize recursive programs. 
In that work, we showed that the entire PSB1 benchmark can be solved by one of four different recursion schemes: catamorphism, accumulation, anamorphism, and hylomorphism. 
It also describes preliminary experiments with catamorphism showing that, for the problems that are solvable with this scheme, the use of scaffolding improved the success rate when compared to HOTGP. 
The results were promising and are evidence that, once the choice of which recursion scheme is made, the synthesis process can be simplified as the evolution searches for simpler trees in a more constrained search space.


\section{Origami Program Synthesis}
\label{sec:methods}



Current Origami implementation follows a Koza-style Genetic Programming~\cite{koza1992genetic} (tree representation) with the main differences from traditional approaches being the introduction of immutable nodes, ensuring a certain recursion scheme, and the type-safety of the genetic operators (as in~\cite{hotgp}).

The implementation is based on \emph{patterns}, which are used to represent different recursion schemes.
A pattern is composed of immutable nodes and a set of evolvable slots that, when replaced with expressions, can be evaluated.
The immutable nodes describe the main definition of the recursion scheme (see \autoref{sec:patterns}) and are fixed once we choose the scheme, while the evolvable slots represent the inner mechanisms that need to be synthetized to correspond to the expected behavior described by the dataset.
These slots have a well-defined output type (inferred from the problem description), and a well-defined set of bindings to which the expression has access.

In this work, we focus on the 6 different patterns that comprehend the minimal set required to solve PSB1.
Naturally, the Origami framework is not limited to these patterns, and more could be introduced as needed.
The following section introduces each of the 6 patterns in technical detail. 
In this text, due to space constraints, we assume the reader has a basic understanding of recursion schemes~\cite{meijer1991functional}, how they can be used to solve PSB1~\cite{origami}, as well as the Haskell language notation, which is similar to the ML notation. 

\subsection{Patterns}
\label{sec:patterns}

\newcommand{\slot}[1]{\textcolor{blue}%
{\code{\textit{\underline{slot\textsubscript{#1}}}}}}

\newcommand{\argVal}[1]{%
\code{arg{\textsubscript{#1}}}%
}

\newcommand{\argType}[1]{%
\code{i{\textsubscript{#1}}}%
}

\newcommand{\scopeitem}[2]{
\item \code{#1}~::~\code{#2}
}

\newenvironment{scope}
{%
    with scope \{\begin{itemize*}[
    itemjoin={;},
    label={} 
    ]
}
{%
    \end{itemize*}%
    \}%
}

\newcommand{\argScope}{
    \item \argVal{0}~::~\argType{0}
    $\ldots$
    \argVal{n}~::~\argType{n}
}

\subsubsection{NoScheme}

This is the simplest pattern in Origami, as it does not employ any recursion at all.
It is represented by the following code:

\begin{minted}[escapeinside=@@]{haskell}
f @\argVal{0}@ ... @\argVal{n}@ = @\slot{1}@
\end{minted}

This pattern has just a single slot, which has all the arguments in scope and returns a value of the same type as the output of the program.
Its main use is to accommodate for problems that do not require any recursion.

\subsubsection{Catamorphism over Indexed List}

This pattern captures the most common recursion scheme observed in PSB1, and arguably in practical scenarios as well, \ie folding a list from the right.
This pattern is commonly used in Haskell as \code{foldr}. In Meijer-notation~\cite{meijer1991functional}, this would be represented by the banana brackets $\llparenthesis b, \oplus \rrparenthesis$, where $b$ is the initial value and $\oplus$ is the combining function.
In the context of Origami, it is represented by the following code:

\begin{minted}[escapeinside=@@]{haskell}
f @\argVal{0}@ ... @\argVal{n}@ = cata alg @\argVal{0}@ where
      alg INil = @\slot{1}@
      alg (ICons i x acc) = @\slot{2}@
\end{minted}

In a problem with arguments of type \code{\argType{0} $\ldots$ \argType{n}} and of output type \code{o}, where \code{\argType{0}~$\equiv$~[e]}%
\footnote{In this text, the notation \code{\argType{0}~$\equiv$~[e]} is a restriction such that \code{\argType{0}} can be decomposed into the type \code{[e]}, which is the type of a list with elements of some type \code{e}.}%
, this pattern's slots are typed as follows:

\begin{itemize}
    \item \code{\slot{1}~::~o}, with nothing in scope;
    \item \code{\slot{2}~::~o},
    \begin{scope}
        \scopeitem{i}{Int}
        \scopeitem{x}{e}
        \scopeitem{acc}{o}
        \argScope
    \end{scope}.
\end{itemize}

The catamorphism solutions presented in~\cite{origami} alternately used regular and Indexed Lists.
In an effort to minimize the number of different patterns to be considered, we chose to only have the Indexed List variation of the catamorphisms, as it is more general than a regular List (\ie, we can make it indexed by replacing \code{[]} with \code{INil}, and \code{x:xs} with \code{ICons index x xs}).
For the remaining schemes we employ a regular list since it is enough to solve the problem (as shown by the canonical solutions presented in~\cite{origami}).

In an effort to minimize the number of different patterns to be considered, we chose to only have the Indexed List variation of this scheme, as it is more general than a regular List.
For brevity, this will be referred to as simply \emph{Cata} in the remainder of this paper.

\subsubsection{Curried Catamorphism over Indexed List}
This pattern captures a common variation of the Catamorphism, and can be represented by the following code:

\begin{minted}[escapeinside=@@]{haskell}
f @\argVal{0}@ @\argVal{1}@ = cata alg @\argVal{0}@ @\argVal{1}@ where
      alg INil = \ys -> @\slot{1}@
      alg (ICons i x f) = \ys -> @\slot{2}@
\end{minted}

As a problem of type \code{\argType{0}~->~\argType{1}~->~o} can also be seen in its curried form as \code{\argType{0}~->~(\argType{1}~->~o)}, we can employ Catamorphism to accumulate a \emph{function} over the first argument, and then apply this function to the second argument.
This is useful when we need to apply a Catamorphism over the zip of two lists~\cite{origami}.

In a problem with arguments of type \code{\argType{0}, \argType{1}}\footnote{It is worth noting that, differently from other patterns, this one can only be applied to problems with exactly two arguments.} and of output type \code{o}, where \code{\argType{0}~$\equiv$~[e]}, this pattern's slots are typed as follows:

\begin{itemize}
    \item \code{\slot{1}~::~o},
    \begin{scope}
        \scopeitem{ys}{\argType{1}}
    \end{scope};
    
    \item \code{\slot{2}~::~o}, 
    \begin{scope}
        \scopeitem{i}{Int}
        \scopeitem{x}{e}
        \scopeitem{f}{\argType{1}~->~o}
        \scopeitem{ys}{\argType{1}}
    \end{scope}.
\end{itemize}

For brevity, this will be referred to as simply \emph{CurriedCata} in the rest of this paper.

\subsubsection{Anamorphism to a List}
This pattern is commonly used in Haskell as \code{unfold}, which is used to generate a list. In Meijer-notation~\cite{meijer1991functional}, this would be represented by the concave lenses $\lbparen g, p\rbparen$
where $g$ is the generator function, and $p$ is the predicate. In the context of Origami, it can be represented by the following code:

\begin{minted}[escapeinside=@@]{haskell}
f @\argVal{0}@ ... @\argVal{n}@ = ana coalg @\slot{1}@ where
      coalg seed = if @\slot{2}@ then [] 
                   else @\slot{3}@ : @\slot{4}@
\end{minted}

In a problem with arguments of type \code{\argType{0} $\ldots$ \argType{n}} and of output type \code{o}, where \code{o~$\equiv$~[e]}, this pattern's slots are typed as follows:

\begin{itemize}
    \item \code{\slot{1}~::~\argType{0}}, 
    \begin{scope}
        \argScope
    \end{scope};
    
    \item \code{\slot{2}~::~Bool}, 
    \begin{scope}
        \scopeitem{seed}{\argType{0}}
        \argScope
    \end{scope};
    
    \item \code{\slot{3}~::~e},
    \begin{scope}
        \scopeitem{seed}{\argType{0}}
        \argScope 
    \end{scope};

    \item \code{\slot{4}~::~\argType{0}},
    \begin{scope}
        \scopeitem{seed}{\argType{0}}
    \end{scope}.
\end{itemize}

Note that while we do not enforce \code{\argVal{0}} to be used in \slot{1}, it must be of the same type as \code{\argVal{0}}, as all of the solutions for PSB1 respected this constraint.
For brevity, this will be referred to as simply \emph{Ana} in the rest of this paper.

\subsubsection{Accumulation over a List}
This pattern captures using an accumulation strategy before using a \code{foldr}, and can be represented by the following code:

\begin{minted}[escapeinside=@@]{haskell}
f @\argVal{0}@ ... @\argVal{n}@ = accu st alg @\argVal{0}@ @\slot{1}@
      where
        st [] s = []
        st (x : xs) s = x : (xs, @\slot{2}@)
        alg [] s = @\slot{3}@
        alg (x : acc) s = @\slot{4}@
\end{minted}

In a problem with arguments of type \code{\argType{0} $\ldots$ \argType{n}} and of output type \code{o}, where \code{\argType{0}~$\equiv$~[e]}, \textbf{and given a type \code{a}}, this pattern's slots are typed as follows:

\begin{itemize}
    \item \code{\slot{1}~::~a}, 
    \begin{scope}\argScope\end{scope};
    
    \item \code{\slot{2}~::~a},
    \begin{scope}
        \scopeitem{x}{e}
        \scopeitem{xs}{[e]}
        \scopeitem{s}{a}
        \argScope
    \end{scope};
    
    \item \code{\slot{3}~::~o},
    \begin{scope}
        \scopeitem{s}{a}
        \argScope
    \end{scope};

    \item \code{\slot{4}~::~o},
    \begin{scope}
        \scopeitem{x}{e}
        \scopeitem{acc}{o}
        \scopeitem{s}{a}
        \argScope
    \end{scope}.
\end{itemize}

Notice that this is the first pattern whose types are not fully determined by the type of the arguments and the expected output type: the accumulator type \code{a}.
In the context of this paper, types such as this one will be referred to as \emph{unbound types}.
To keep the implementation simple, we assume unbound types are known and provided by the user.
Even though that might not be the case in a practical scenario, it is not possible to try all types as there is an infinite number of them.
Properly exploring different types is an interesting challenge that warrants dedicated research.
For brevity, this pattern will be referred to as simply \emph{Accu} in the rest of this paper.

\subsubsection{Hylomorphism through a List}
This pattern captures an Anamorphism followed by a Catamorphism, such as applying \code{foldr} to the result of \code{unfold}, in Haskell.
In Meijer-notation~\cite{meijer1991functional}, this would be represented by the envelopes $\left\llbracket (c, \oplus), (g, p)\right\rrbracket$.
In Origami, it is represented by the following code:

\begin{minted}[escapeinside=@@]{haskell}
f @\argVal{0}@ ... @\argVal{n}@ = hylo alg coalg @\argVal{0}@ where
      coalg seed = if @\slot{1}@ then [] else @\slot{2}@ : @\slot{3}@
      alg [] = @\slot{4}@
      alg (x : acc) = @\slot{5}@
\end{minted}

In a problem with arguments of type \code{\argType{0} $\ldots$ \argType{n}} and of output type \code{o}, \textbf{and given a type \code{a}}, this pattern's slots are typed as follows:

\begin{itemize}
    \item \code{\slot{1}~::~Bool},
    \begin{scope}
        \scopeitem{seed}{\argType{0}}
        \argScope
    \end{scope};
    
    \item \code{\slot{2}~::~a},
    \begin{scope}
        \scopeitem{seed}{\argType{0}}
        \argScope 
    \end{scope};
    
    \item \code{\slot{3}~::~\argType{0}},
    \begin{scope}
        \scopeitem{seed}{\argType{0}}
        \argScope 
    \end{scope};

    \item \code{\slot{4}~::~o}, with nothing in scope;
    
    \item \code{\slot{5}~::~o}, 
    \begin{scope}
        \scopeitem{x}{a}
        \scopeitem{acc}{o}
        \argScope 
    \end{scope}.
\end{itemize}

This is also a pattern that contains an unbound type: the intermediary list has elements of type \code{a}.
For brevity, this pattern will be referred to as simply \emph{Hylo} in the rest of this paper.

\subsection{Genetic Programming}

Origami synthesizes the evolvable slots using a Genetic Programming (GP)~\cite{koza1992genetic} algorithm. 
Since each one of the recursive patterns requires more than a single slot, we represent each solution as a collection of programs represented as expression trees.
Each element of this collection corresponds to one of the slots.

The GP starts with an initial random population of $1\,000$ individuals, and iterates by applying either crossover to a pair of parents, or mutation to a single parent, generating $1\,000$ new individuals in total.
The entire population is replaced by the offspring population.

The initial population is generated using a ramped half-and-half where half of the individuals are generated using the full method and half using the grow method.
The maximum depth for each method varies between $1$ and $5$.
The parental selection is performed using a tournament selection of size $10$.

Following a simple GP algorithm, in Origami the mutation randomly selects one of the evolvable slots, then picks one point in the tree at random to be replaced by a new subtree generated at random using the grow method, with a maximum depth of $5 - d_\mathit{current}$.
Crossover also starts by picking one of the slots at random, then performing one of these two actions with equal probability: 
i) swap the entire slot of one parent with the same slot of the other parent; 
ii) swap two subtrees of the same output type from each parent.



\subsection{Grammar}

In HOTGP (our previous work~\cite{hotgp}), the choice of grammar was focused on providing a minimal set of operations that would enable functional programming, with special attention to higher-order functions.

With Origami, however, our main focus is assessing whether the inclusion of Recursion Schemes is beneficial to Program Synthesis.
To achieve this goal, we must enforce that recursion can only happen by selecting the appropriate pattern.
Therefore, our grammar was designed not to use any implicitly recursive functions, such as \code{map}, \code{filter}, \code{sum}, and \code{product}, being aware that by doing it we might remove shortcuts and potentially make the synthesis of certain problems harder.
Changes were also made in an effort to more closely match the set of operations used by other methods, with special attention to the grammar used in recent implementations of PushGP~\cite{helmuth2018program}. As a result, Origami has a larger set of operations than HOTGP.
The full grammar is presented in~\autoref{tab:grammar}.

\begin{table}[p!]
    \caption{The complete set of operations available for Origami. Each dataset only had access to the operations that involved its allowed types according to~\cite{psb1}.}
    \label{tab:grammar}
    \centering
    \begin{tabular}{p{.5\linewidth}|p{.5\linewidth}}
    \toprule
        Operations & Types \\
        \midrule
\code{addInt, subInt, multInt, divInt, quotInt, modInt, remInt, minInt, maxInt} & \code{Int -> Int -> Int} \\
\code{absInt, succInt, predInt } & \code{Int -> Int} \\
\code{addFloat, subFloat, multFloat, divFloat, minFloat, maxFloat} & \code{Float -> Float -> Float} \\
\code{absFloat, sqrt, sin, cos, succFloat, predFloat} & \code{Float -> Float} \\
\code{fromIntegral} & \code{Int -> Float} \\
\code{floor, ceiling, round} & \code{Float -> Int} \\
\code{ltInt, gtInt, gteInt, lteInt} & \code{Int -> Int -> Bool} \\
\code{ltFloat, gtFloat, gteFloat, lteFloat} & \code{Float -> Float -> Bool} \\
\midrule
\code{and, or} & \code{Bool -> Bool -> Bool} \\
\code{not} & \code{Bool -> Bool} \\
\code{if} & \code{Bool -> a -> a -> a} \\
\code{eq, neq} & \code{a -> a -> Bool} \\
\midrule
\code{showInt} & \code{Int -> [Char]} \\
\code{showFloat} & \code{Float -> [Char]} \\
\code{showBool} & \code{Bool -> [Char]} \\
\code{showChar} & \code{Char -> [Char]} \\
\code{charToInt} & \code{Char -> Int} \\
\code{intToChar} & \code{Int -> Char} \\
\code{isLetter, isSpace, isDigit} & \code{Char -> Bool} \\
\midrule
\code{length} & \code{[a] -> Int} \\
\code{cons, snoc} & \code{a -> [a] -> [a]} \\
\code{mappend} & \code{[a] -> [a] -> [a]} \\
\code{elem} & \code{a -> [a] -> Bool} \\
\code{delete} & \code{a -> [a] -> [a]} \\
\code{null} & \code{[a] -> Bool} \\
\code{head, last} & \code{[a] -> a} \\
\code{tail, init} & \code{[a] -> [a]} \\
\code{zip} & \code{[a] -> [b] -> [(a, b)]} \\
\code{replicate} & \code{Int -> a -> [a]} \\
\code{enumFromThenTo} & \code{Int -> Int -> Int -> [Int]} \\
\code{reverse} & \code{[a] -> [a]} \\
\code{splitAt} & \code{Int -> [a] -> ([a], [a])} \\
\code{intercalate} & \code{[a] -> [a] -> [a]} \\
\midrule
\code{fst} & \code{(a, b) -> a} \\
\code{snd} & \code{(a, b) -> b} \\
\code{mkPair} & \code{a -> b -> (a, b)} \\
\midrule
\code{apply} & \code{(a -> b) -> a -> b} \\
\midrule
\code{singleton} & \code{a -> b -> Map a b} \\
\code{insert} & \code{a -> b -> Map a b -> Map a b} \\
\code{insertWith} & \code{((b, b) -> b) -> a -> b -> Map a b -> Map a b} \\
\code{fromList} & \code{[(a, b)] -> Map a b} \\
         \bottomrule
    \end{tabular}
\end{table}

Once an execution is finished, the champion's slots are refined using the same procedure that was used in HOTGP~\cite{hotgp}.
To refine a tree, we pick the root node and check if replacing it with any of its children leads to a correctly-typed solution with an equal or better fitness.
If so, we replace it with the best child; otherwise, we keep the original node.
This process continues recursively, traversing the tree and greedily replacing nodes with their children when needed.
This procedure aims to apply Occam's Razor and generate a simpler and more general solution~\cite{helmuth2017improving}, making sure the fitness in the training dataset is never worse.

\section{Experimental Results}
\label{sec:results}

To evaluate our approach we conducted experiments to perform an automatic search for different patterns in the PSB1~\cite{psb1} context.
For each of the $29$ datasets, we sequentially tried each pattern in increasing order of complexity:

\begin{enumerate}
    \item NoScheme;
    \item Cata, if \code{\argVal{0}} is a list;
    \item CurriedCata, if the problem has two arguments and \code{\argVal{0}} is a list;
    \item Ana, if the return type is a list;
    \item Accu, if \code{\argVal{0}} is a list;
    \item Hylo.
\end{enumerate}

For each dataset, $30$ seeds of each pattern were executed, and we only advanced to the next applicable pattern if none of the seeds succeeded in finding a solution (\ie the success rate was $0\%$).
Each seed followed the instructions provided by PSB1, using the recommended number of training and test instances, and included the fixed edge cases in the training data, as well as using the fitness functions described in~\cite{psb1}.

Note that we deliberately placed the patterns with unbound types at the end of the sequence.
Therefore, the unbound type in both Accu and Hylo is only decided after all other schemes have failed.
For most benchmarks that got to this point, we chose the type that was known to be correct according to the solutions presented in~\cite{origami} (from now on referred to as \emph{canonical solutions}).
For the cases in which the canonical solutions did not use Accu or Hylo,  we chose a reasonable  type as needed. These choices are summarized in~\autoref{tab:unbound_types}.

\newcommand{\bestResult}{\underline}
\newcommand{\canonical}{\color{blue}}
\begin{table}[t]
    \centering
    \caption{The chosen types for the unbound types in Accu and Hylo. The type is colored in blue when the decision was guided by the canonical solution.}
    \label{tab:unbound_types}
    \begin{tabular}{lll}
    \toprule
       \textbf{Dataset} & Accu & Hylo  \\
       \midrule
        checksum & \canonical\code{Int} & \code{Int} \\
        collatz-numbers & -- & \canonical\code{Int} \\
        digits & -- & \code{Int} \\
        pig-latin & -- & \code{[Char]} \\
        string-differences & \code{Int} & \code{(Char, Char)} \\
        sum-of-squares & -- & \canonical\code{Int} \\
        vector-average & \canonical\code{(Float, Int)} & \code{Int} \\
        wallis-pi & -- & \canonical\code{Float} \\
        word-stats & \canonical\code{((Int, Int), (Int, Int))} & \code{[Char]} \\
        x-word-lines & \canonical\code{Int} & \code{[Char]} \\
         \bottomrule
    \end{tabular}
\end{table}

The maximum tree depth was set to $5$ for each slot, and the crossover rate was empirically set to $50\%$. 
We allowed a maximum of $300\,000$ evaluations with an early stop whenever the algorithm finds a perfectly accurate solution according to the training data.
For patterns in which termination is not guaranteed, namely Ana and Hylo, a maximum number of iterations was imposed (empirically set to $10\,000$).
We also encountered an issue specific to CurriedCata, where Origami was synthesizing solutions with the slot \code{alg (Cons i x f) = \textbackslash ys -> f (f ys)}, essentially creating a ``fork bomb''.
To prevent this issue from happening, we introduced a maximum execution budget, which will terminate the evaluation of the entire individual when a single iteration of a slot applied more than $10\,000$ operations.

\autoref{tab:results_by_pattern} shows the percentage of executions in which Origami was able to synthesize a solution that completely solved the test set (\ie success rate).
Origami found a solution for all of the problems that were canonically solved by NoScheme as well as Cata.
Surprisingly, it was also able to synthesize a solution for 
``for-loop-index'' by using NoScheme, even though the canonical solution used Ana, and for ``grade'' by using Cata, even though the canonical solution used CurriedCata.
We nonetheless ran both of these problems with their canonical patterns and discovered Origami was also able to synthesize solutions, albeit less often.
Moreover, Origami was able to find the solutions for $3$ out of the $4$ canonical CurriedCata problems, and $2$ out of the $3$ Ana problems.
Accu and Hylo, however, appear to be the most difficult patterns to synthesize, as no solution for problems that canonically involve these patterns was found.

\begin{table}[t!]
\caption{
Percentual success rates obtained by Origami for each pattern in each dataset. 
The ``Best'' column shows the highest success rate obtained for that dataset across all patterns, which is also underlined.
We also show in blue the pattern of the canonical solution.
}
\label{tab:results_by_pattern}
    \centering

\begin{tabular}{lrrrrrr|r}
\toprule
\textbf{Dataset} & NoScheme & Cata & CurriedCata & Ana & Accu & Hylo & \textbf{Best} \\
\midrule
checksum & 0 & 0 & -- & -- & \canonical{0} & 0 & 0 \\
collatz-numbers & 0 & -- & -- & -- & -- & \canonical{0} & 0 \\
compare-string-lengths & \canonical{\bestResult{90}} & -- & -- & -- & -- & -- & 90 \\
count-odds & 0 & \canonical{\bestResult{40}} & -- & -- & -- & -- & {40} \\
digits & 0 & -- & -- & \canonical{0} & -- & 0 & 0 \\
double-letters & 0 & \canonical{\bestResult{3}} & -- & -- & -- & -- & {3} \\
even-squares & 0 & -- & -- & \canonical\bestResult{3} & -- & -- & {3} \\
for-loop-index & \bestResult{90} & -- & -- & \canonical{67} & -- & -- & {90} \\
grade & 0 & \bestResult{100} & \canonical{10} & -- & -- & -- & {100} \\
last-index-of-zero & 0 & \canonical\bestResult{70} & -- & -- & -- & -- & {70} \\
median & \canonical\bestResult{97} & -- & -- & -- & -- & -- & {97} \\
mirror-image & \canonical\bestResult{93} & -- & -- & -- & -- & -- & {93} \\
negative-to-zero & 0 & \canonical\bestResult{87} & -- & -- & -- & -- & {87} \\
number-io & \canonical\bestResult{100} & -- & -- & -- & -- & -- & {100} \\
pig-latin & 0 & \canonical0 & -- & -- & -- & 0 & 0 \\
replace-space-with-newline & 0 & \canonical\bestResult{3} & -- & -- & -- & -- & {3} \\
scrabble-score & 0 & \canonical\bestResult{100} & -- & -- & -- & -- & {100} \\
small-or-large & \canonical\bestResult{53} & -- & -- & -- & -- & -- & {53} \\
smallest & \canonical\bestResult{100} & -- & -- & -- & -- & -- & {100} \\
string-differences & 0 & 0 & \canonical{0} & -- & 0 & 0 & 0 \\
string-lengths-backwards & 0 & \canonical\bestResult{97} & -- & -- & -- & -- & {97} \\
sum-of-squares & 0 & -- & -- & -- & -- & \canonical0 & 0 \\
super-anagrams & 0 & 0 & \canonical\bestResult{73} & -- & -- & -- & {73} \\
syllables & 0 & \canonical\bestResult{7} & -- & -- & -- & -- & {7} \\
vector-average & 0 & 0 & -- & -- & \canonical{0} & 0 & 0 \\
vectors-summed & 0 & 0 & \canonical\bestResult{20} & -- & -- & -- & {20} \\
wallis-pi & 0 & -- & -- & -- & -- & \canonical{0} & 0 \\
word-stats & 0 & 0 & -- & -- & \canonical{0} & 0 & 0 \\
x-word-lines & 0 & 0 & 0 & -- & \canonical{0} & 0 & 0 \\
\bottomrule
\end{tabular}
\end{table}

Considering the $4$ canonical Accu problems, \emph{checksum} and \emph{word-stats} are historically hard, with few methods ever finding a solution.
The same can be said for Hylo in the \emph{wallis-pi} and \emph{collatz-numbers} problems.

In \emph{vector-average}, the canonical solution involved using Accu to compute both the sum and the count as a pair in the \code{st} slots, and using the \code{alg} slots to perform the division as a post-processing step, finally obtaining the average.
The solution that got closer to the intended result was the following:
\begin{minted}[escapeinside=@@]{haskell}
accu st alg @\argVal{0}@ (last @\argVal{0}@, length @\argVal{0}@)
  where
    st [] s = []
    st (x : xs) s = x : (xs, s)

    alg [] s = min 0 (last @\argVal{0}@)
    alg (x : acc) s = acc + ((max (x - acc) x) / (snd s))
\end{minted}

Origami took a different approach from the canonical solution, by storing the length of the input in the second element of the tuple while having no use for the first element.
The \code{st} section had no other purpose than to transmit this pre-processing step to the \code{alg} section.
This solution got a perfect score during training but failed in testing for certain cases.
If we were to replace \code{min 0 (last \argVal{0})} by \code{0} and \code{max (x - acc) x} by \code{x}, then this solution would be correct.


The Hylo solution for \emph{sum-of-squares} employed \code{coalg} to generate a list of all the numbers from $0$ to \argVal{0}, and then used \code{alg} to square each number and accumulate the sum.
Even though this was the simplest Hylo solution, as Hylo has $5$ different slots, it has an increased search space in relation to other patterns, which seems to be a big challenge for the algorithm.

\autoref{tab:vs_hotgp} compares Origami's results to HOTGP's.
There was a substantial increase ($>30$) in the success rate in $6$ problems.
In the $17$ problems where the absolute difference is $<30$, we highlight \textit{syllables},\textit{ double-letters} and \textit{even-squares} problems, as those were problems for which HOTGP was not able to synthesize a solution, whereas Origami was successful at least once.
The two problems with a more noticeable decrease are \textit{replace-space-with-newline} and \textit{vector-average}.
These can be explained by the change in grammar between the two algorithms, as HOTGP's solutions were arguably simpler due to having \code{map} and \code{filter} for \textit{replace-space-with-newline} and \code{sum} for \textit{vector-average}.
In a practical scenario, the inclusion of these functions would likely lead to a correct solution but, as previously noted, removing them was a conscious decision to enable the proper assessment of the impact of Recursion Schemes in PS.
It would also allow for composite solutions, such as using Ana with a \code{map} inside instead of relying on Hylo to find the entire pattern, which might be easier to synthesize.

\begin{table}[t!]
    \caption{Origami's success rates compared to HOTGP's. The $\Delta$ column shows the relative success rate of Origami with respect to HOTGP. Problems not solved by either approach are omitted for clarity.}
    \label{tab:vs_hotgp}
    \centering
\begin{tabular}{lrrr}
\toprule
Dataset & Origami & HOTGP & $\Delta$ \\
 \midrule
scrabble-score & 100 & 0 & 100 \\
mirror-image & 93 & 1 & 92 \\
super-anagrams & 73 & 0 & 73 \\
last-index-of-zero & 70 & 0 & 70 \\
grade & 100 & 37 & 63 \\
for-loop-index & 90 & 59 & 31 \\
string-lengths-backwards & 97 & 89 & 8 \\
syllables & 7 & 0 & 7 \\
double-letters & 3 & 0 & 3 \\
even-squares & 3 & 0 & 3 \\
smallest & 100 & 100 & 0 \\
number-io & 100 & 100 & 0 \\
sum-of-squares & 0 & 1 & -1 \\
median & 97 & 99 & -2 \\
small-or-large & 53 & 59 & -6 \\
compare-string-lengths & 90 & 100 & -10 \\
count-odds & 40 & 50 & -10 \\
negative-to-zero & 87 & 100 & -13 \\
vectors-summed & 20 & 37 & -17 \\
replace-space-with-newline & 3 & 38 & -35 \\
vector-average & 0 & 80 & -80 \\
\bottomrule
\end{tabular}
\end{table}

To position Origami in the landscape of the best solutions currently found in the literature, we compare the obtained results against those obtained by PushGP~\cite{psb1}, Grammar-Guided Genetic Programming (G3P)~\cite{g3p}, and the extended grammar version of G3P (here called G3P+)~\cite{g3pe}, and some recently proposed methods such as Code Building Genetic Programming (CBGP)~\cite{pantridge2022functional}, and G3P with Haskell and Python grammars (G3Phs and G3Ppy)~\cite{garrow2022functional}. 
The results are reported in \autoref{tab:vs_all}. 
In this table, the ``--'' means the authors did not test their algorithm for that specific benchmark.

\begin{table}[p]
    \caption{Success rate for each dataset. The best values for each problem are highlighted.
    The ``\# of Best'' row shows the amount of problems in which each method obtained the best result.
    The final $5$ rows show the amount of problems in which each method obtained a success rate above a certain threshold.
    }
    \label{tab:vs_all}
    \centering
    \renewcommand{\arraystretch}{1.1}
\begin{tabularx}{\textwidth}{X|rr|rrrr|rrrr}
 \toprule
 \scriptsize	\textbf{Dataset} &  \scriptsize	Origami & \scriptsize	HOTGP &\scriptsize	 DSLS & \scriptsize	UMAD & \scriptsize	PushGP & \scriptsize	CBGP & \scriptsize	G3P & \scriptsize	G3P+ & \scriptsize	G3Phs & \scriptsize	G3Ppy \\
 \midrule
checksum & 0 & 0 & \bestResult{18} & 5 & 0 & -- & 0 & 0 & -- & -- \\
collatz-numbers & 0 & -- & -- & -- & 0 & -- & 0 & 0 & -- & -- \\
compare-string-lengths & 90 & \bestResult{100} & 51 & 42 & 7 & 22 & 2 & 0 & 5 & 0 \\
count-odds & 40 & \bestResult{50} & 11 & 12 & 8 & 0 & 12 & 3 & -- & -- \\
digits & 0 & 0 & \bestResult{28} & 11 & 7 & 0 & 0 & 0 & -- & -- \\
double-letters & 3 & 0 & \bestResult{50} & 20 & 6 & -- & 0 & 0 & -- & -- \\
even-squares & \bestResult{3} & 0 & 2 & 0 & 2 & -- & 1 & 0 & -- & -- \\
for-loop-index & \bestResult{90} & 59 & 5 & 1 & 1 & 0 & 8 & 6 & -- & -- \\
grade & \bestResult{100} & 37 & 2 & 0 & 4 & -- & 31 & 31 & -- & -- \\
last-index-of-zero & \bestResult{70} & 0 & 65 & 56 & 21 & 10 & 22 & 44 & 0 & 2 \\
median & 97 & \bestResult{99} & 69 & 48 & 45 & 98 & 79 & 59 & 96 & 21 \\
mirror-image & 93 & 1 & 99 & \bestResult{100} & 78 & \bestResult{100} & 0 & 25 & -- & -- \\
negative-to-zero & 87 & \bestResult{100} & 82 & 82 & 45 & 99 & 63 & 13 & 0 & 66 \\
number-io & \bestResult{100} & \bestResult{100} & 99 & \bestResult{100} & 98 & \bestResult{100} & 94 & 83 & 99 & \bestResult{100} \\
pig-latin & 0 & 0 & 0 & 0 & 0 & -- & 0 & \bestResult{3} & -- & -- \\
replace-space-with-newline & 3 & 38 & \bestResult{100} & 87 & 51 & 0 & 0 & 16 & -- & -- \\
scrabble-score & \bestResult{100} & 0 & 31 & 20 & 2 & -- & 2 & 1 & -- & -- \\
small-or-large & 53 & \bestResult{59} & 22 & 4 & 5 & 0 & 7 & 9 & 4 & 0 \\
smallest & \bestResult{100} & \bestResult{100} & 98 & \bestResult{100} & 81 & \bestResult{100} & 94 & 73 & \bestResult{100} & 89 \\
string-differences & 0 & -- & -- & -- & 0 & -- & -- & -- & -- & -- \\
string-lengths-backwards & \bestResult{97} & 89 & 95 & 86 & 66 & -- & 68 & 18 & 0 & 34 \\
sum-of-squares & 0 & 1 & 25 & \bestResult{26} & 6 & -- & 3 & 5 & -- & -- \\
super-anagrams & \bestResult{73} & 0 & 4 & 0 & 0 & -- & 21 & 0 & 5 & 38 \\
syllables & 7 & 0 & \bestResult{64} & 48 & 18 & -- & 0 & 39 & -- & -- \\
vector-average & 0 & 80 & \bestResult{97} & 92 & 16 & 88 & 5 & 0 & 4 & 0 \\
vectors-summed & 20 & 37 & 21 & 9 & 1 & \bestResult{100} & 91 & 21 & 68 & 0 \\
wallis-pi & 0 & -- & -- & -- & 0 & -- & 0 & 0 & -- & -- \\
word-stats & 0 & -- & -- & -- & 0 & -- & 0 & 0 & -- & -- \\
x-word-lines & 0 & 0 & \bestResult{91} & 59 & 8 & -- & 0 & 0 & -- & -- \\
\midrule
\textbf{\# of Best} & \bestResult{9} & 7 & 7 & 4 & 0 & 4 & 0 & 1 & 1 & 1 \\
\midrule
$\mathbf{= 100\%}$ & \bestResult{4} & \bestResult{4} & 1 & 3 & 0 & \bestResult{4} & 0 & 0 & 1 & 1 \\
$\mathbf{\geq 75\%}$ & \bestResult{10} & 7 & 8 & 7 & 3 & 7 & 4 & 1 & 3 & 2 \\
$\mathbf{\geq 50\%}$ & \bestResult{13} & 10 & \bestResult{13} & 9 & 5 & 7 & 6 & 3 & 4 & 3 \\
$\mathbf{\geq 25\%}$ & 14 & 13 & \bestResult{16} & 13 & 7 & 7 & 7 & 7 & 4 & 5 \\
$\mathbf{> 0\%}$ & 19 & 15 & \bestResult{24} & 21 & 22 & 9 & 17 & 17 & 8 & 7 \\
\bottomrule
\end{tabularx}
\end{table}

Origami has the best results in $9$ of the problems, which is the highest across all methods.
It also has the highest number of problems solved with $100\%$, $\geq75\%$, and $\geq50\%$, and is second-place in $\geq25\%$.
When we consider problems to which Origami found at least one solution, we note that it outperforms HOTGP, CBGP, and all the G3P variations, placing Origami at the fourth place overall. 
It is also worth noticing that Origami outperforms HOTGP in both the number of best results and amount of problems above all thresholds, which demonstrates it is a substantial improvement over HOTGP.

\section{Conclusion}
\label{sec:conclusion}

This work is the first full implementation of Origami, a GP algorithm proposed in~\cite{origami}, and builds on our previous work, HOTGP~\cite{hotgp}.
Origami's main differential is the use of Recursion Schemes, well-known constructs in functional programming that enable recursive algorithms to be defined in a unified manner.
The main motivation for using these in the PS context is enabling recursive programs to be synthesized in a controlled manner, without sacrificing expressiveness.

We evaluate our approach in the 29 problems in the PSB1 dataset, which is known to be solvable by just a handful of Recursion Schemes.
In general, Origami performs better than other methods, synthesizing the correct solution more often than others in $9$ problems, which is more than any other algorithm.
It was able to obtain the highest count of problems with success rate $=100\%$, $\geq75\%$ and $\geq50\%$.
These experimental results suggest that using Recursion Schemes to guide the search is a promising research avenue.
Currently, the main challenge of Origami appears to be dealing with harder Recursion Schemes, such as Accumulation and Hylomorphism. 
Different evolutionary mechanisms, such as other selection methods and mutation/crossover operators, should be evaluated in this context to understand if they can positively impact the search process.

In future work, a scheme to deal with unbound types will need to be developed. 
Some schemes, such as Accumulation and Hylomorphism, are not fully defined by the input and output types of the problem and need extra type guidance (on the unbound types) to properly define the program.
In this implementation, this information is provided by the user as a hyperparameter; but in a practical scenario, this might be undesirable.
Ideally, the search for the ``correct'' unbound type should be incorporated into the search algorithm.



\bibliographystyle{splncs04}
\bibliography{bib}
\end{document}